\documentstyle[prl,eqsecnum,aps]{revtex}

\begin{document}
\author{Guoyou Huang  \footnote{E-mail address: hgyphysics@yahoo.com.cn}}
\address{Beihai Cambridge Research Center, 107 South Wenming St. Beihai, Guangxi
Province. 536000, P.R.CHINA}
\date{\today}
\title{On Structure and History of Space-time with Variable
Speed of Light} \maketitle

\begin{abstract}
We apply the variable speed of light into general relativity in
order to solve the problems we met in the standard cosmology.
We're surprised to find that, the results from the general
relativity in cosmology are exactly the same as those we got from
Newtonian dynamics. The relation between the Newtonian dynamics
and the relativistic dynamics can be demonstrated with the
variable speed of light. With this approach, some problems in the
standard cosmology such as the flatness problem and the horizon
problem doesn't arise any more. All the cosmological results and
the physical results are reasonable and natural. There are no any
difficulties in the standard cosmology.

\end{abstract}
\section{ Introductions}

There're two serious problems in the standard cosmology, flatness
problem and horizon problem [1]. The cosmology has gone too far
before these two problems were solved. In an early paper [2], I
introduced the variable speed of light to the Newtonian dynamics,
and successfully used it to solve these two problems. Here in this
paper, I further introduce the same speed of light to the general
relativity in order to solve the problems in cosmology under the
general relativity. Surprisingly, we found that, the results from
the general relativity are exactly the same as those from the
Newtonian dynamics. These encourage us to believe that the
relations between Newtonian dynamics and relativity can be
demonstrated. And the speed of light is changed with the age of
universe. But this variable of speed of light doesn't violate the
invariable principle of the speed of light under relativity.
Because the former is the change in the magnitude of the speed,
and the latter is the independence of speed to the observers.

\section{ Basic Structure of Cosmos}

With the spherical coordinates system, the metric of cosmos has
the simple form as

\begin{equation}
\label{eq1} ds^2 = - e^\nu dt^2 + e^\mu dr^2 + r^2(d\theta ^2 +
\sin ^2\theta d\varphi ^2)
\end{equation}

The Schwarzschild's exterior solution of the cosmos system is

\begin{equation}
\label{eq2} ds^2 = - \left( {1 - \frac{2GM}{r}} \right)dt^2 +
\left( {1 - \frac{2GM}{r}} \right)^{ - 1}dr^2 + r^2\left( {d\theta
^2 + \sin ^2\theta .d\varphi ^2} \right)
\end{equation}

M and r are mass and radius of the cosmos respectively.

In the cosmos system, the gravitational constant G and light speed
C have the following relations [2]

\begin{equation}
\label{eq3} G = ZC
\end{equation}

\begin{equation}
\label{eq4} C = \frac{2ZM}{R}
\end{equation}

R is the radius of cosmos system. Z is a constant, $Z =
2.224\times 10^{ - 19}m^2s^{ - 1}kg^{ - 1}$.

From (\ref{eq2}), we can see that the surface of the cosmos is a
special interface. The characteristic of the metric tensor is

\begin{equation}
\label{eq5}
\begin{array}{l}
 g_{00} = 0 £ g^{00} = \infty \\
 g_{11} = \infty £ g^{11} = 0 \\
 \end{array}
\end{equation}

The observer outside this interface can never get any information
from inside the interface. So, the whole cosmos system is a giant
black hole to the outside world. But for the variable of speed of
light with matter distribution inside the cosmos, there is no any
special area in the cosmos [2]. It's to say, motion of matter in
any direction inside the cosmos is possible, but matter needs an
infinite time to reach this special interface. The Black Hole
Paradox about the whole universe system was solved.

The Schwarzschild's interior solution is

\begin{equation}
\label{eq6} e^{\nu / 2} = \frac{3}{2}(1 -
\frac{2GM}{R})^{\frac{1}{2}} - \frac{1}{2}(1 -
\frac{2GMr^2}{R^3})^{\frac{1}{2}}
\end{equation}

\begin{equation}
\label{eq7} e^{ - \mu } = 1 - \frac{2GMr^2}{R^3} r \le R
\end{equation}

The distribution of pressure inside the cosmos system is given as
[3]

\begin{equation}
\label{eq8} p(r) = \rho \frac{(1 -
\frac{2GMr^3}{R^3})^{\frac{1}{2}} - (1 -
\frac{2GM}{R})^{\frac{1}{2}}}{3(1 - \frac{2GM}{R})^{\frac{1}{2}} -
(1 - \frac{2GMr^2}{R^3})^{\frac{1}{2}}}
\end{equation}

Notice that the medium of the cosmos is extremely relativistic,
the thermal motion speed of most of the matter in cosmos included
the photons and neutrinos, and other fields is exactly the speed
of light, so the pressure inside the cosmos is isotropy. And

\begin{equation}
\label{eq9} p = \rho / 3
\end{equation}

\section{ the Standard Model of Cosmos}

The standard model of the cosmos was described with the
Robertson-Walker metric [3]

\begin{equation}
\label{eq10} ds^2 = - dt^2 + R^2(t)(\frac{dr^2}{1 - kr^2} +
r^2d\theta ^2 + r^2\sin ^2\theta .d\varphi ^2)
\end{equation}

The energy-momentum tensor for idea fluids of the cosmos is given
by

\begin{equation}
\label{eq11} T^{\mu \nu } = (p + \rho )U^\mu U^\nu + pg^{\mu \nu }
\end{equation}

$U^\mu = (1,0,0,0).$Apply (\ref{eq10}) and (\ref{eq11}) to the
following field equation

\begin{equation}
\label{eq12} R_{\mu \nu } = - 8\pi G(T_{\mu \nu } -
\frac{1}{2}g_{\mu \nu } T)
\end{equation}

We get the time-time component as

\begin{equation}
\label{eq13} 3\ddot {R} = - 4\pi G(\rho + 3p)R
\end{equation}

\noindent and space-space component as

\begin{equation}
\label{eq14} R\ddot {R} + \dot {R}^2 + 2k = 4\pi G(\rho - p)R^2
\end{equation}

Then we get from the last two equations the following differential
equation

\begin{equation}
\label{eq15} \dot {R}^2 + k = \frac{8\pi G}{3}\rho R^2
\end{equation}

The conservation law of energy and momentum requires the energy
momentum tensor in (\ref{eq11}) satisfies with

\begin{equation}
\label{eq16} T_{ ;\nu }^{\mu \nu } = 0
\end{equation}

Its time component gives

\begin{equation}
\label{eq17} \frac{d\rho }{dt} + \frac{3\dot {R}}{R}(\rho + p) = 0
\end{equation}

Together with (\ref{eq9}), we have all the equations needed to get
the space, pressure and density relation R(t), p(t) and $\rho
$(t).

\section{ the Curvature of Cosmos}
\label{subsubsec:mylabel2}

We reasonable assume that the interface of cosmos expands with the
speed of light. The Hubble parameter can be defined as

\begin{equation}
\label{eq18} H = \frac{\dot {R}}{R} = \frac{C}{R} =
\frac{2ZM}{R^2}
\end{equation}

The basic equation (\ref{eq15}) can be changed into

\begin{equation}
\label{eq19} \rho = \rho _c + \frac{3}{8\pi G}\frac{k}{R^2}
\end{equation}

\begin{equation}
\label{eq20} \rho _c \equiv \frac{3H^2}{8\pi G}
\end{equation}

From (\ref{eq3}) and (\ref{eq4}), we get

\begin{equation}
\label{eq21} \rho _c \equiv \frac{3H^2}{8\pi G} = \frac{3M}{4\pi
R^3} = \rho
\end{equation}

(\ref{eq19}) gives

\begin{equation}
\label{eq22} k \equiv 0
\end{equation}

It's to say, in any period of the cosmos, the space is always
flat. We can prove this conclusion by the followings.

The matter and entropy densities of the cosmos from the Planck
formula are given as

\begin{equation}
\label{eq23} \rho = \frac{\pi ^2}{30}NT^4
\end{equation}

\begin{equation}
\label{eq24} s = \frac{2\pi ^2}{45}NT^3
\end{equation}

N is the total degree of freedom. The particle physics gives $N =
10^2$

The expansion of cosmos is entropy conservative. It's to say

\begin{equation}
\label{eq25} S = sR^3 = const.
\end{equation}

(\ref{eq24}) and (\ref{eq25}) show that

\begin{equation}
\label{eq26} TR = const.
\end{equation}

(\ref{eq19}) can be changed into

\begin{equation}
\label{eq27} 1 - \frac{\rho _c }{\rho } = \frac{3k}{8\pi
G}\frac{1}{\rho R^2}
\end{equation}

Together with (\ref{eq23}), (\ref{eq24}) and (\ref{eq25}),
(\ref{eq27}) can be further changed into

\begin{equation}
\label{eq28} 1 - \frac{\rho _c }{\rho } = \frac{3k}{8\pi
G}(\frac{160}{3\pi ^2N})^{\frac{1}{2}}S^{ - \frac{2}{3}}T^{ - 2}
\end{equation}

We use the entropy from the background radiations to substitute
the entropy of cosmos

\begin{equation}
\label{eq29} S \approx S_\gamma = s_\gamma R^3 = \frac{2\pi
^2}{45}T_\gamma ^3 R^3
\end{equation}

Substitute the data of today's cosmos to (\ref{eq29}), we get

\begin{equation}
\label{eq30} S \ge 10^{87}
\end{equation}

In the early cosmos when $t = 10^{ - 20}s$, $T = 10^5Gev$, $G =
6.71\times 10^{ - 21}Gev^{ - 2}$, so

\begin{equation}
\label{eq31} 1 - \frac{\rho _c }{\rho } \le 10^{ - 61}
\end{equation}

This proves that the space is strictly flat even in the early
cosmos when the gravitation is much stronger. This solved the
flatness problem.

\section{ the Space-time and Light Speed of Cosmos}
\label{subsubsec:mylabel3}

The decelerate parameter of the cosmos is defined as

\begin{equation}
\label{eq32} q = - \frac{\ddot {R}R}{\dot {R}^2}
\end{equation}

The expansion of the cosmos with the light speed is extremely
relativistic, so

\begin{equation}
\label{eq33} p = \rho / 3
\end{equation}

From (\ref{eq13}) we obtain

\begin{equation}
\label{eq34} q = \frac{\rho }{\rho _c }
\end{equation}

From (\ref{eq13}) and (\ref{eq15}), we obtain

\begin{equation}
\label{eq35} \frac{k}{R^2} = H^2(q - 1)
\end{equation}

Equation (\ref{eq17}) has an equivalent form as

\begin{equation}
\label{eq36} \frac{d}{dR}(\rho R^3) = - 3pR^2
\end{equation}

From equation (\ref{eq33}) and (\ref{eq36}) we obtain

\begin{equation}
\label{eq37} \rho R^4 = const.
\end{equation}

Rewrite the basic equation (\ref{eq15}) with (\ref{eq35}) and
(\ref{eq37}), we obtain

\begin{equation}
\label{eq38} \dot {R}^2 = \frac{8\pi G\rho _0 R_0 ^4}{3R^2} -
R_0^2 H_0^2 (q_0 - 1)
\end{equation}

In (\ref{eq38}), all the parameters are those for today's cosmos.
Notice that $\rho _0 $, $H_0 $and $q_0 $ have the relations shown
in equation (\ref{eq20}) and (\ref{eq34}). Substitute $\rho _0 $
from (\ref{eq38}), we obtain

\begin{equation}
\label{eq39} \left( {\frac{\dot {R}}{R}} \right)^2 = H_0^2 \left(
{1 - q_0 + q_0 \frac{R_0^2 }{R^2}} \right)
\end{equation}

Let $X = R / R_0 $, then $dR / R_0 = dX$, (\ref{eq39}) can be
changed into

\begin{equation}
\label{eq40} dt = \frac{1}{H_0 }(1 - q_0 + q_0 X^{ - 2})^{ -
\frac{1}{2}}dX
\end{equation}

From (\ref{eq21}) and (\ref{eq34}), we can see that $q_0 = 1$,
equation (\ref{eq40}) can be changed into

\begin{equation}
\label{eq41} dt = \frac{X}{H_0 }dX
\end{equation}

\begin{equation}
\label{eq42} t = \frac{X^2}{2H_0 }\left| {{\begin{array}{*{20}c}
 {R / R_0 } \hfill \\
 0 \hfill \\
\end{array} }} \right.
\end{equation}

Notice that $H_0 = 2ZM / R_0^2 $, so

\begin{equation}
\label{eq43} t = \frac{R^2}{4ZM}
\end{equation}

\noindent or

\begin{equation}
\label{eq44} R(t) = 2Z^{\frac{1}{2}}M^{\frac{1}{2}}t^{\frac{1}{2}}
\end{equation}

From (\ref{eq18}), the Hubble parameter is given as

\begin{equation}
\label{eq45} H = \frac{1}{2}t^{ - 1}
\end{equation}

And the speed of light is given as

\begin{equation}
\label{eq46} C = Z^{\frac{1}{2}}M^{\frac{1}{2}}t^{ - \frac{1}{2}}
\end{equation}

From above we can see that the Hubble parameter and the speed of
light are changed with the age of universe.

\section{ the Thermodynamics of Cosmos}
\label{subsubsec:mylabel4}

The space is strictly flat, so the basic equation (\ref{eq15}) can
be rewrite as

\begin{equation}
\label{eq47} (\frac{\dot {R}}{R})^2 = \frac{8\pi G}{3}\rho
\end{equation}

Apply (\ref{eq23}) and (\ref{eq26}) to (\ref{eq47}), the
temperature equation of cosmos can be obtained as

\begin{equation}
\label{eq48} \frac{dT}{dt} = - (\frac{4\pi
^3NG}{45})^{\frac{1}{2}}T^3
\end{equation}

\begin{equation}
\label{eq49} T(t) = (\frac{45}{16\pi ^3NG})^{\frac{1}{4}}t^{ -
\frac{1}{2}}
\end{equation}

The density equation can be obtained from (\ref{eq23}) and
(\ref{eq49}) as

\begin{equation}
\label{eq50} \rho (t) = \frac{3t^{ - 2}}{32\pi G}
\end{equation}

\section{ the Horizon of Cosmos}
\label{subsubsec:mylabel5}

The motion of photons satisfies $ds = 0$, this can be obtained
from equation (\ref{eq10}) as

\begin{equation}
\label{eq51} \frac{dr}{\sqrt {1 - kr^2} } = - \frac{dt}{R(t)}
\end{equation}

The horizon radius of the cosmos is defined as

\begin{equation}
\label{eq52} \int_0^{r(H)} {\frac{dr}{\sqrt {1 - kr^2} }} =
\int_0^t {\frac{dt}{R(t)}}
\end{equation}

The horizon radius is in fact the instantaneous distance $l_{H(t)}
$ from $r = 0$ to $r = R_{(H)} $

\begin{equation}
\label{eq53} l_{H(t)} \equiv R(t)\int_0^{r(H)} {\frac{dr}{\sqrt {1
- kr^2} }} = R(t)\int_0^t {\frac{dt'}{R(t')}}
\end{equation}

(\ref{eq44}) shows that R is in direct proportion to
$t^{\frac{1}{2}}$. Solve the equation (\ref{eq53}) we get

\begin{equation}
\label{eq54} l_{H(t)} = 2t
\end{equation}

Notice the nature unit system and speed of light (\ref{eq46}), the
horizon of the cosmos (\ref{eq54}) is exactly the radius of the
cosmos (\ref{eq46}) in any moment of the cosmos. We can also prove
this conclusion with the observations by the followings.

In the moment of $10^{ - 20}s$, the horizon of cosmos is about
$10^8m$ from (\ref{eq54}). And the temperature of the cosmos is
about $10^5Gev$ from (\ref{eq49}). Now let's calculate the scale
of our observed 10 billion light-year ($10^{26}m)$ cosmos with a
temperature of about 2.8K. From (\ref{eq26}) , we obtain

\begin{equation}
\label{eq55} R = \frac{T_0 R_0 }{T} = \frac{2.8K\times
10^{26}m}{10^5\times 10^{13}K} \approx 10^8m
\end{equation}

It's exactly the horizon at that moment. So, there is no any
abnormity in the whole history of cosmos. This solved the horizon
problem.

\section{ the History of Cosmos}
\label{subsubsec:mylabel6}

All the parameters of cosmos in any moment can be calculated
accurately with equation (\ref{eq44}),(\ref{eq45}),(\ref{eq46}),
(\ref{eq49}) and (\ref{eq50}), included the radius R, light
velocity C, gravitational constant G, temperature T and density
$\rho $ of the cosmos. Table.8.1 shows these parameters in four
important epochs in the history of cosmos.

\textbf{Table 8.1 Four Important Epochs in History of Cosmos}

\begin{table}
\begin{tabular}
{|p{50pt}|p{45pt}|p{64pt}|p{55pt}|p{206pt}|} \hline t (s)& $R
(m)$& \textit{$\rho $(}$kgm^{ - 3})$& T&
Events \\
\hline $10^{ - 20}$& $10^8$& $10^{30}$& $10^5Gev$&
\textbf{Big Nucleus} epoch. \\
\hline $10^{ - 14}$& $10^{10}$& $10^{22}$& 1\textit{Gev}&
\textbf{Big Atom }epoch. Hadrons formed. \\
\hline $10^{ - 4}$& $10^{16}$& $10^5$& 0.1\textit{mev}&
\textbf{Big Star} epoch. Nuclei formed. \\
\hline $10^{10}$& $10^{22}$& $10^{ - 15}$& 1\textit{ev}&
\textbf{Big Galaxy} epoch. Atom formed. \\
\hline
\end{tabular}
\label{tab1}
\end{table}

All the results calculated in table 8.1 are accordant with the
conditions for events in every period of cosmos. It enables us to
study the early cosmos with quantitative analysis.

\section{ the Abundance of Helium}
\label{subsubsec:mylabel7}

We can see from table 8.1 that in the early cosmos of $10^{ -
4}$s, the density is about $10^5kgm^{ - 3}$, and the temperature
is about 0.1mev. These are exactly the conditions for the
combination of helium in stars. The helium in cosmos is mainly
formed in this period.

The combination of helium began from deuteron ${ }_1^2 D$. The
binding energy of deuteron is 2.2mev. The combination of deuteron
happened in the time when the temperature of cosmos dropped to
0.1mev, or $10^9$K. In the stars, when the temperature reaches
$10^7$K and the density reaches $10^5kgm^{ - 3}$, the combination
of helium can be started. In the Big Star, the temperature was
higher and the interaction was much stronger. So the combination
completed very thoroughly. All of the neutrons were combined into
helium with protons.

In a definite condition, the neutrons and protons are in a
reversible equilibrium. They can be transformed into each other
with the following reactions

\begin{equation}
\label{eq56}
\begin{array}{l}
 p + l^ - = n + v_l \\
 n + l^ + = p + \tilde {v}_l \\
 \end{array}
\end{equation}

The numerical density of neutrons $N_n $ and protons $N_p $ were
given by the Boltzmann formula as

\begin{equation}
\label{eq57} \frac{N_n }{N_p } = e^{ - \Delta m / T}
\end{equation}

$\Delta m = m_n - m_p = 1.29mev$is the difference of the mass
between neutron and proton.

From (\ref{eq57}) we can see that, when the temperature of the
cosmos dropped to 0.1mev for the combination of deuterons, $N_n $
is apparently less than $N_p $. From the weak interaction theory
in particle physics, we know that when the density dropped to
$10^{10}kgm^{ - 3}$, the transformation between neutrons and
protons stopped. The density of neutrons was fixed at this density
limit. From the above equations we can see that this density was
happened in about $10^{ - 5}s$, when the radius was
about$10^{14}m$. From (\ref{eq50}) we can see that the
corresponded temperature is 0.8mev. It was just before the
combination of deuteron and helium. When the temperature dropped
to 0.1mev. The combination of helium was started. All the neutrons
were combined into helium. The remained protons are the hydrogen.
At that time, the density of helium and hydrogen has the following
relations

\begin{equation}
\label{eq58}
\begin{array}{l}
 N_{He} = \frac{1}{2}N_n \\
 N_H = N_p - N_n \\
 \end{array}
\end{equation}

The abundance of helium was so defined as

\begin{equation}
\label{eq59} Y_{He} = \frac{2N_{He} }{N_H + 4N_{He} } = \frac{2}{1
+ N_p / N_n } = \frac{2}{1 + e^{\Delta m / T}}
\end{equation}

Apply the temperature 0.8mev at which the transformation between
neutrons and protons stopped to (\ref{eq59}) yields

\begin{equation}
\label{eq60} Y_{He} = 0.28
\end{equation}

\section{ Conclusions}

The critical density is exactly the density of cosmos. The space
of cosmos is strictly flat anytime and anywhere in the cosmos. The
horizon of cosmos is exactly the radius of cosmos. The physical
events in any period of cosmos are accordant with the matter
conditions calculated by equations presented in this paper. There
is no any abnormity in any period or area of the cosmos. All the
cosmological results and the physical results are reasonable and
natural. There are no any difficulties in the standard cosmology
under the general relativity. An experiment is necessary to test
the variable speed of light. It's not difficult, but it still
needs someone to finish it. It must be significant to approach the
unification of Newtonian and relativistic dynamics.

\end{document}